\documentclass[10pt,conference]{IEEEtran}

\usepackage{cite}
\usepackage{csquotes}
\usepackage{amsmath,amssymb,amsfonts}
\usepackage{algorithmic}
\usepackage{graphicx}
\usepackage{textcomp}
\usepackage[colorlinks=false,hidelinks]{hyperref}
\usepackage[capitalise,nameinlink,noabbrev]{cleveref}
\usepackage{xcolor}
\usepackage{threeparttable}
\usepackage{multirow}
\usepackage{array}
\usepackage{makecell}
\usepackage{listings}
\usepackage{pifont}
\newcommand{\cmark}{\ding{51}}
\newcommand{\xmark}{\ding{55}}

\usepackage{diagbox}

\usepackage[square,numbers,sort&compress]{natbib}

\usepackage[framemethod=TikZ]{mdframed}
\usepackage{xpatch}

\usepackage[square,numbers,sort&compress]{natbib}

\usepackage[framemethod=TikZ]{mdframed}
\usepackage{xpatch}

\makeatletter
\xpatchcmd{\endmdframed}
  {\aftergroup\endmdf@trivlist\color@endgroup}
  {\endmdf@trivlist\color@endgroup\@doendpe}
  {}{}
\makeatother

\begin{document}

\title{Qunicorn: A Middleware for the Unified Execution Across Heterogeneous Quantum Cloud Offerings}

\author{
\IEEEauthorblockN{Benjamin Weder, Johanna Barzen, Martin Beisel, \\Fabian Bühler, Daniel Georg, Frank Leymann, Lavinia Stiliadou} 
\IEEEauthorblockA{\textit{University of Stuttgart, Institute of Architecture of Application Systems, Universitatsstraße 38, Stuttgart, Germany}\\
\textit{\{firstname.lastname\}@iaas.uni-stuttgart.de}}
}

\maketitle

\begin{abstract}
Quantum computers are available via a variety of different quantum cloud offerings.
These offerings are heterogeneous and differ in features, such as pricing models or types of access to quantum computers.
Furthermore, quantum circuits can be implemented using different quantum programming languages, which are typically only supported by a small subset of quantum cloud offerings.
As a consequence, using a specific quantum programming language for implementing the application at hand can limit the set of compatible quantum cloud offerings and cause a vendor lock-in.
Therefore, selecting a suitable quantum cloud offering and a corresponding quantum programming language requires knowledge about their features.
In this paper, we (i)~analyze the available quantum cloud offerings and extract their features.
Moreover, we (ii)~introduce the architecture for a unification middleware that facilitates accessing quantum computers available via different quantum cloud offerings by automatically translating between various quantum circuit and result formats.
To showcase the practical feasibility of our approach, we (iii)~present a prototypical implementation and validate it for three exemplary application scenarios.
\end{abstract}


\section{Introduction}
\label{sec:introduction}
\noindent
Quantum computers from different providers are publicly available via the cloud, e.g., IBM, Google, IonQ, Xanadu, and Rigetti~\cite{Leymann2020_QuantumCloud,LaRose2019,nguyen2024_quantumcloudreview}.
Compared to traditional hardware, quantum computers are expected to provide advantages for certain problems, such as speed-ups, higher precision, and improved energy efficiency~\cite{Barzen2021_QuantumComputing,nielsen2002quantum}.
Therefore, applications from various areas, e.g., machine learning, chemistry, and scientific simulations, can benefit from utilizing quantum computing~\cite{national2019quantum,cao2018qcchemistry,havlivcek2019supervised}.
\looseness=-1

However, the current software and hardware landscape in the quantum computing domain is very heterogeneous~\cite{weder2022SEDevLifecycle, Vietz2021_QuantumSoftwareEngineeringChallenges}.
For example, to implement a quantum circuit a variety of technologies can be used:
(i)~Quantum programming languages, such as \textit{Q\#}, (ii)~quantum assembly languages, such as \textit{OpenQASM}, and (iii)~quantum \textit{Software Development Kits~(SDKs)} that are embedded into classical programming languages, such as \textit{Qiskit} or \textit{Forest} in Python~\cite{LaRose2019,fingerhuth2018open, Vietz2021_OnDecisionSupport, cross2017open}.
As quantum cloud offerings typically only support a limited set of these technologies, implementing a quantum circuit using \mbox{a specific technology often leads to a vendor lock-in~\cite{Vietz2021_QuantumSoftwareEngineeringChallenges,weder2022SEDevLifecycle}.}

The quantum computers available via the various quantum cloud offerings differ in their characteristics, such as the number of provided qubits or the gate error rates~\cite{tannu2019, Weder2021_QuantumProvenance}.
However, these characteristics affect the executability of quantum circuits and the quality of the execution results~\cite{Salm2020_NISQAnalyzer}.
As quantum computers are typically used together with classical computers and solve specific sub-problems within a larger hybrid quantum application, the problem instance solved by a quantum circuit often results from previous computations~\cite{Leymann2020_QuantumAlgorithmsBitterTurth}.
Depending on the quantum circuit size, another quantum computer might be better suited for execution~\cite{Salm2020_NISQAnalyzer}.
Thus, the selection of a suitable quantum computer must be performed during runtime of the application~\cite{Weder2021_QuantumWorkflowHardwareSelection}.
However, if the circuit format is not supported by the quantum cloud offering providing the selected \mbox{quantum computer, the quantum circuit has to be translated.}\looseness=-1

Moreover, not only the underlying quantum computers but also the quantum cloud offerings vary in their functionality, types of access, and pricing models~\cite{Vietz2021_QuantumSoftwareEngineeringChallenges}.
For example, some quantum cloud offerings only support the execution of quantum circuits using a queue while others enable exclusive reservations.
However, due to the heterogeneity of the available quantum cloud offerings and the lack of a comprehensive evaluation of their supported features, the selection of a suitable quantum cloud offering is a complex and time-consuming task.

In this paper, we (i)~analyze existing quantum cloud offerings to identify their features, e.g., the supported types of access and the used quantum circuit formats.
This analysis supports users in identifying a suitable quantum cloud offering for their application.
Furthermore, we (ii)~present an architecture and a prototype for a middleware that facilitates accessing quantum computers available via different quantum cloud offerings.
This middleware unifies the access and automatically transforms the quantum circuit and result formats.
Moreover, it also provides additional functionalities, e.g., cutting large quantum circuits or unified observability when utilizing multiple heterogeneous quantum cloud offerings.
To validate our approach, we (iii)~utilize the unification middleware to execute multiple quantum circuits within three typical application scenarios in the quantum computing domain.\looseness=-1

The remainder of this paper is as follows:
\Cref{sec:fundamentals} introduces the fundamentals and discusses our problem statement.
In \Cref{sec:api}, we present our comparison of the features of different quantum cloud offerings.
\cref{sec:qunicorn} shows the architecture for the middleware unifying the access to various quantum cloud offerings. 
Furthermore, its practicality is verified by a prototypical implementation that is utilized for implementing three exemplary application scenarios.
\Cref{sec:discussion} discusses the approach and its limitations.
Finally, \Cref{sec:rw} presents related work, and \Cref{sec:conclusion} concludes the paper.

\vspace{-2px}
\section{Fundamentals \& Problem Statement}
\label{sec:fundamentals}
\noindent
\vspace{-1px}
In this section, we introduce fundamentals about quantum applications and how to access quantum computers of various providers, followed by a discussion of our problem statement.\looseness=-1

\vspace{-2px}
\subsection{Hybrid Quantum Applications \& Quantum Cloud Offerings}
\noindent
\vspace{-1px}
Quantum applications are typically hybrid since they involve the execution of quantum algorithms, but also rely on classical programs, e.g., to interact with the user or visualize results~\cite{Weder2021_OrchestrationsInSuperposition,weder2022SEDevLifecycle}.
Additionally, quantum algorithms are often hybrid as they require classical pre-processing of data and post-processing of results or even classical processing as part of the overall algorithm~\cite{Leymann2020_QuantumAlgorithmsBitterTurth,Barzen2022_Shor}.
For example, so-called \textit{variational quantum algorithms} use parameterized quantum circuits and optimize these parameters classically in multiple iterations~\cite{cerezo2021variational,harrow2021low}.\looseness=-1

Quantum computers are accessed via heterogeneous quantum cloud offerings, e.g., IBMQ and AWS Braket~\cite{Leymann2020_QuantumCloud, LaRose2019}.
For example, they differ in their access methods.
Typically quantum computers can be accessed through a queue-based model, where a job is submitted to the queue, waits for execution, and then results are retrieved~\cite{LaRose2019}.
To reduce waiting time, certain offerings enable reserving exclusive time slots on the quantum computer~\cite{Vietz2021_QuantumSoftwareEngineeringChallenges}.
Some quantum cloud offerings also support executing hybrid programs comprising classical and quantum logic and optimize their execution~\cite{Weder2022_QuantumWorkflowRewrite}.
Another example is the availability of quantum computers with different capabilities, e.g., the number of provided qubits or error rates~\cite{tannu2019, Weder2021_QuantumProvenance}.

\vspace{-2px}
\subsection{Problem Statement \& Research Questions}
\vspace{-1px}
\noindent
With the growing number of quantum cloud offerings and their heterogeneity regarding the supported features, such as pricing models or access methods, selecting a suitable offering for an application is a complex and time-consuming task.
To make a decision, one must systematically identify and evaluate the key features of each publicly accessible quantum cloud offering.
Therefore, the first Research Question~(RQ) is as follows:

\begin{mdframed}[skipabove=4pt,skipbelow=0.5pt,innerbottommargin=2.5px,innertopmargin=3.5px\baselineskip, innerleftmargin=8px, innerrightmargin=8px]
\hspace{-5px}
\textit{\textbf{RQ 1:} ``What are relevant features of quantum cloud offerings, and how do the various quantum cloud offerings differ regarding the support of these features?''}
\end{mdframed}

\noindent
The most suitable quantum cloud offering for a quantum application at hand might vary over time, e.g., due to changes in the pricing model or the supported quantum computers.
However, switching between quantum cloud offerings is difficult due to differences in supported programming languages and SDKs.
Adapting an existing quantum application for a different quantum cloud offering may require rewriting it utilizing a different technology that is supported by the respective quantum cloud offering.
Translating an entire quantum application is very time-consuming, particularly when developers are unfamiliar with the required technologies.
This hinders interoperability and complicates the usage of quantum computers across \mbox{different quantum cloud offerings.
Hence, our second RQ is:}

\begin{mdframed}[skipabove=4pt,skipbelow=0.5pt,innerbottommargin=2.5px,innertopmargin=3.5px\baselineskip, innerleftmargin=8px, innerrightmargin=8px]
\hspace{-5px}
\textit{\textbf{RQ 2:} ``How to execute quantum circuits independently of the heterogeneity of quantum cloud offerings regarding their supported quantum circuit and result formats?''}
\end{mdframed}

\noindent
To address these RQs, we analyzed different quantum cloud offerings regarding their supported feature set, which is presented in the next section.
Furthermore, we realized a unification middleware to enable using various quantum cloud offerings utilizing a single, unified API, as described in \cref{sec:qunicorn}.\looseness=-1

\section{Comparison of Quantum Cloud Offerings}
\label{sec:api}
\noindent
In this section, we present our comparison of different quantum cloud offerings.
We first describe the data sources used and the search strategy applied to identify available quantum cloud offerings.
Subsequently, the evaluation criteria are discussed, and the comparison results are presented in detail.
To ensure the validation and replication of all steps in our comparison, the collected data is available on GitHub~\cite{USTUTT_replication}.

\begin{figure}[tb!]
	\centering
	\includegraphics[width=\linewidth]{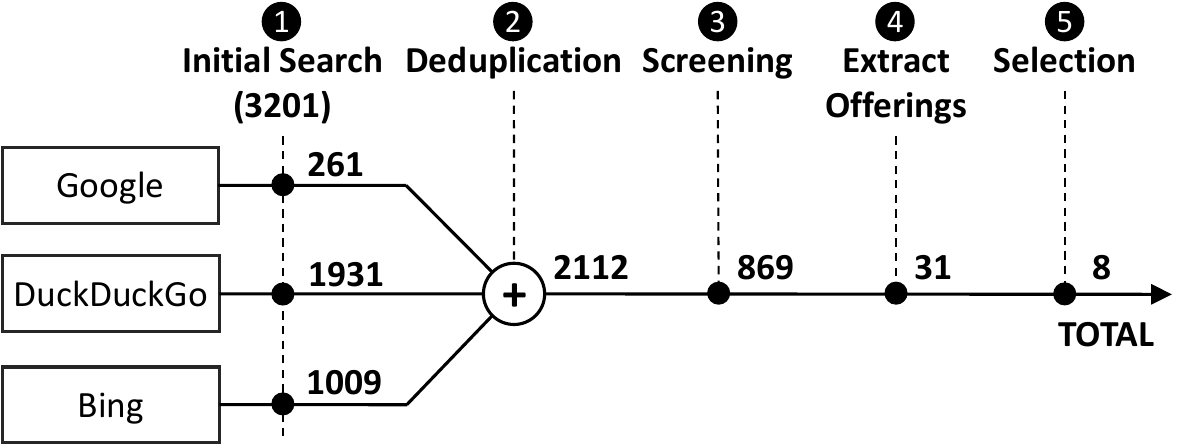}
	\caption{Overview of the search and selection process}
	\label{fig:searchAndSelectionStrategy}
\end{figure}

\subsection{Data Sources and Search Strategy}
\noindent
As quantum cloud offerings typically provide a website with information regarding their functionalities, we followed the guidelines for performing a gray literature review to identify the available quantum cloud offerings~\cite{paez2017gray,garousi2019guidelines}.
An overview of our applied search and selection strategy is shown in~\Cref{fig:searchAndSelectionStrategy}.

\noindent
\textbf{Initial Search:}
In the initial step of our search process, we define a search string that aims to identify a large set of websites that present or discuss quantum cloud offerings: \looseness=-1

\begin{lstlisting}[columns=fullflexible, label=lst:query,caption=Query string used for the initial search]
        quantum computing AND (cloud OR platform 
           OR service OR offering OR company)
\end{lstlisting}

\noindent
To retrieve a wide range of search results, we used the search string with three different search engines, namely Google, DuckDuckGo, and Bing.
The initial search yielded a total of 3201 entries, with the majority originating from DuckDuckGo. \looseness=-1

\begin{table*}[htb!]
    \centering
    \caption{An overview of the quantum cloud offerings and the evaluation criteria.}
    \vspace{-3px}
    \label{tab:services}
    \renewcommand{\arraystretch}{1.32} 
    \begin{tabular}{ >{\centering}m{2.5cm} | c | c | c | c | c | c | c | c | c |}
        \diagbox[font=\small, height=0.9cm, innerwidth=2cm]{Criteria}{Offering}& 
        \makecell{\textsc{Alice \&} \\ \textsc{Bob}}  &  
        \makecell{\textsc{AWS} \\ \textsc{Braket}}  &
        \makecell{\textsc{Azure} \\ \textsc{Quantum}} &
        \makecell{\textsc{IBMQ}} &
        \makecell{\textsc{IonQ}  \\ \textsc{Quantum} \\ \textsc{Cloud}} &
        \makecell{\textsc{qBraid} \\ \textsc{Quantum} \\ \textsc{Platform}} &
        \makecell{\textsc{Quandela} \\ \textsc{Cloud}} &
        \makecell{\textsc{Quantum} \\ \textsc{Inspire}}\\
        
    \hline
    \vspace{-6px}
            \makecell{\textbf{Access Models} \\ \textit{Queue}
            \\ \textit{Prioritized Queue}
            \\ \textit{Exclusive Time Slots}} & 
             \makecell{ \\ \checkmark \\ \xmark \\ \checkmark} &
             \makecell{ \\ \checkmark \\ \checkmark \\ \checkmark} &
             \makecell{ \\ \checkmark \\ \checkmark \\ \xmark} &
             \makecell{ \\ \checkmark \\ \checkmark \\ \checkmark} &
             \makecell{ \\ \checkmark \\ \xmark \\ \checkmark} &
             \makecell{ \\ \checkmark \\ \xmark \\ \xmark} &
             \makecell{ \\ \checkmark \\ \checkmark \\ \checkmark} &
             \makecell{ \\ \checkmark \\ \xmark \\ \xmark} \\
    \hline
            \makecell{\textbf{Batch Processing}} & \checkmark & \checkmark & \xmark & \checkmark & \checkmark & \xmark & \checkmark & \xmark \\
    \hline

            \makecell{\textbf{Circuit Cutting}} & \xmark  & \xmark & \xmark & \xmark & \xmark & \xmark & \xmark & \xmark \\
    \hline
    \vspace{-3px}
            \makecell{\textbf{Error Handling} \\ \textit{Error Mitigation} \\ \textit{Error Correction}} &  
            \makecell{ \\ \xmark \\ \xmark }  
            & \makecell{ \\ \checkmark \\ \xmark }
            & \makecell{ \\ \checkmark \\ \xmark } 
            & \makecell{ \\ \checkmark \\ \xmark } 
            & \makecell{ \\ \checkmark \\ \xmark }
            & \makecell{ \\ \checkmark \\ \xmark } 
            & \makecell{ \\ \checkmark \\ \xmark } 
            & \makecell{ \\ \xmark \\ \xmark } \\
    \hline

            \makecell{\textbf{\textit{Hybrid Runtime}}} & \xmark  & \checkmark & \checkmark & \checkmark & \xmark & \xmark & \xmark & \xmark \\
    \hline
    \vspace{-3px}
            \makecell{\textbf{\textit{Integrated Dev Env}} \\ \textit{Composer}\\ \textit{Online Code Editor}} & 
            \makecell{ \\ \xmark \\ \xmark } & 
            \makecell{ \\ \xmark \\ \checkmark } &
            \makecell{ \\ \xmark \\ \checkmark } &
            \makecell{ \\ \checkmark \\ \xmark } &
            \makecell{ \\ \xmark \\ \xmark } &
            \makecell{ \\ \checkmark \\ \checkmark } &
            \makecell{ \\ \xmark \\ \checkmark } &
            \makecell{ \\ \xmark \\ \checkmark } \\
    \hline
    \vspace{-8px}
            \makecell{\textbf{\textit{Pricing Models}} \\ \textit{Time-based} \\ \textit{Circuit-based} \\ \textit{Subscription-based}}& 
            \makecell{ \\ \checkmark \\ \xmark \\ \xmark} &
            \makecell{ \\ \checkmark \\ \checkmark \\ \xmark} &
            \makecell{ \\ \checkmark \\ \checkmark \\ \checkmark} &
            \makecell{ \\ \checkmark \\ \xmark \\ \xmark} &
            \makecell{ \\ \xmark \\ \checkmark \\ \xmark} &
            \makecell{ \\ \xmark \\ \checkmark \\ \xmark} &
            \makecell{ \\ \xmark \\ \checkmark \\ \checkmark} &
            \makecell{ \\ \xmark \\ \xmark \\ \xmark} \\
    \hline    
    
            \makecell{\textbf{\textit{Trial Access}}} & \checkmark  & \xmark & \checkmark & \checkmark & \xmark & \xmark & \checkmark & \checkmark\\
    \hline
    \end{tabular}
\end{table*}

\noindent
\textbf{Deduplication:}
In the second step, the initial search results from the different search engines are merged into a single dataset.
Thereby, we removed all duplicate entries based on the websites' URLs, resulting in a total of 2112 websites.

\noindent
\textbf{Screening:}
In the third phase, the list of websites is pruned based on their relevancy for identifying quantum cloud offerings.
Websites are removed if they are not in English or can not be fully accessed.
Furthermore, academic papers and videos are removed from the result set.
To avoid errors during the screening phase, the contents of each website are briefly skimmed by two independent researchers.
After the screening was completed, a total of 869 relevant websites remained.

\noindent
\textbf{Extract Offerings:}
In the fourth phase, we analyzed the content of each website for references to quantum cloud offerings to create an exhaustive list of quantum cloud offerings.
\mbox{In total 31} distinct quantum cloud offerings were identified.

\noindent
\textbf{Selection:}
In the final phase of the search and selection process, the identified quantum cloud offerings were filtered based on a set of inclusion~(\cmark) and exclusion~(\xmark) criteria:\looseness=-1

\begin{itemize}
	\item[\cmark] The documentation and dashboard of the quantum cloud offering are in English and can be fully accessed.
	\item[\cmark] The quantum cloud offering is open to the public and enables the automated creation of new accounts without the need for an application or using a contact form.
    \item[\cmark] The quantum cloud offering is documented such that the available features can be used and evaluated.
    \item[\xmark] The quantum cloud offering only provides access to simulators, not quantum computers.
    \item[\xmark] The quantum cloud offering is marked as a beta, early access, or pre-release version on the website of the offering.\looseness=-1
    \item[\xmark] The quantum cloud offering is a research prototype.
    \item[\xmark] The quantum cloud offering utilizes tokens from other quantum cloud offerings and forwards them for execution instead of, e.g., providing its own pricing model.
    \item[\xmark] The quantum cloud offering only provides access to quantum annealers and not to universal quantum computers.
\end{itemize}

\noindent
The resulting set of entries contains eight quantum cloud offerings.
These offerings are evaluated in detail in the next section.\looseness=-1

\subsection{Evaluation Criteria}
\noindent
The following discusses the evaluation criteria for assessing the quantum cloud offerings.
An overview of the evaluation criteria is shown in the left column of \Cref{tab:services}.
The criteria were identified by analyzing the literature, e.g., \cite{nguyen2024_quantumcloudreview,moguel2024development,Vietz2021_QuantumSoftwareEngineeringChallenges,falkenthal2024planqk,Vietz2021_OnDecisionSupport}.\looseness=-1

\noindent
\textbf{Access Models:}
Quantum cloud offerings utilize different methods to provide access to quantum computers via the cloud.
Generally, there are three access models that range from equal resource sharing to dedicated and exclusive access~\cite{Vietz2021_QuantumSoftwareEngineeringChallenges}:
\textit{Queue}-based access models share the execution time on a quantum computer between multiple users.
When submitted, each quantum circuit is positioned at the end of the queue and executed once it arrives at the front.
The \textit{Prioritized Queue} model can help to improve the efficiency of hybrid quantum applications, especially variational quantum algorithms~\cite{karalekas2020quantum}.
As they frequently alternate between classical and quantum computations, their execution time would significantly increase if they had to go through the entire queue again after every iteration.
Finally, \textit{Exclusive Time Slots} provide dedicated access to a quantum computer for a specific time window.
This type of access is particularly useful when conducting large experiments that require a large number of circuit executions.

\noindent
\textbf{Batch Processing:}
Many application scenarios require executing multiple quantum circuits, e.g., the \textit{Variational Quantum Eigensolver~(VQE)} algorithm~\cite{tilly2022VQE} or the computation of a parameter landscape~\cite{Truger2022_HyperparameterSelection}.
To facilitate the execution of a large number of circuits, some quantum cloud offerings provide batch processing capabilities that enable packaging multiple quantum circuits into a single job.
When the job is executed, all quantum circuits are executed sequentially, and finally, a list containing the execution results for each circuit is returned.\looseness=-1

\noindent
\textbf{Circuit Cutting:}
Circuit cutting techniques are used to break down quantum circuits into smaller sub-circuits~\cite{Peng2019,perlin2021quantum}.
Since these sub-circuits are more shallow and require a smaller number of qubits, circuit cutting enables the execution of circuits whose width exceeds the number of physically available qubits of the quantum computer~\cite{Bechtold2023_CuttingPatterns}.
Furthermore, the smaller circuit size can reduce the negative effect of noise on the execution result~\cite{tang2021cutting}.
To facilitate the utilization of circuit cutting techniques, they can be directly incorporated into quantum cloud offerings so that the techniques can be applied automatically.  \looseness=-1

\noindent
\textbf{Error Handling:}
Current quantum computers are noisy and suffer from various kinds of errors, e.g., gate errors or readout errors~\cite{Salm2020_CriterionExecutingCircuit}.
To reduce the negative impact of these errors, different error-handling techniques have been introduced~\cite{Beisel2022_ErrorHandlingPatterns}: 
\textit{Error Mitigation} techniques aim to reduce the impact of occurring errors while trying to maintain the size of the quantum circuit, e.g., by classical estimating the error-free result based on the quantum computer's error rates~\cite{Beisel2022_ConfigurableReadoutErrorMitigation}.
On the other hand, \textit{Error Correction} methods include error correction codes into the quantum circuit, which drastically increases the circuit size, but enables error-free quantum computations if the error rates of the quantum computer are below a certain threshold~\cite{Knill2001}.
Quantum cloud offerings can facilitate the application of different error-handling methods by enabling their automatic utilization for quantum circuit executions.

\noindent
\textbf{Hybrid Runtime:}
Due to the hybrid nature of quantum applications, their execution typically involves the deployment of classical programs in the cloud or on a local machine~\cite{Weder2021_OrchestrationsInSuperposition}.
However, this complicates the execution of hybrid quantum applications and leads to additional integration challenges~\cite{Beisel2024_ObservabilityForQuantumWorkflows}.
A hybrid runtime provides a hybrid execution environment, enabling to deploy and manage the quantum and classical parts of hybrid quantum applications together~\cite{Vietz2021_QuantumSoftwareEngineeringChallenges}.
Further, these hybrid runtimes can optimize the execution of hybrid quantum applications.
For example, by deploying the classical programs close to the quantum computer to reduce latency when multiple switches between the execution environments are required, e.g., when executing variational quantum algorithms~\cite{Weder2022_QuantumWorkflowRewrite}.

\noindent
\textbf{Integrated Development Environment:}
While the development of large quantum applications is typically done in a local development environment, online development environments are an easy-to-use and accessible option for different scenarios.
For example, a graphical \textit{Circuit Composer} that enables executing the composed circuits can provide great educational value when trying to understand the fundamentals of quantum computing.
Further, practitioners in the field who want to experiment with quantum computing can benefit from an \textit{Online Code Editor} that has all necessary dependencies pre-installed. \looseness=-1

\noindent
\textbf{Pricing Models:}
Quantum cloud offerings use different pricing models to charge users for the utilized resources:
\textit{Time-based} pricing models charge users depending on how long they use a quantum computer. 
For exclusive time slots, the time a quantum computer is reserved is charged as utilized time, even when no circuits are executed.
Quantum cloud offerings using \textit{Circuit-based} pricing models, charge users based on the number of executed quantum circuits. 
Additionally, the size of the circuit may also be incorporated into the pricing calculation.
Finally, \textit{Subscription-based} pricing models require a regular payment, e.g., monthly, and provide a certain amount or unlimited execution time on quantum computers.\looseness=-1

\noindent
\textbf{Trial Access:}
Since accessing quantum computers is currently expensive, the availability of trial periods is important for individuals without the financial support of a company or research institution to familiarize themselves with different quantum cloud offerings.
Trial access must be free of charge and open to the public, i.e., not restricted, e.g., for academic usage.\looseness=-1

\subsection{Results of the Comparison}
\label{subsec:surveyResults}
\noindent
The results of our quantum cloud offering comparison are shown in \cref{tab:services}.
Thereby, a checkmark~(\cmark) indicates that the feature is supported by the quantum cloud offering and suitable documentation mentioning the feature and explaining its usage exists.
In contrast, a cross~(\xmark) means that the feature is either not supported or no documentation was found.

\noindent
\textbf{Access Models:}
All evaluated quantum cloud offerings support queue-based access to their quantum computers.
Furthermore, priority access is provided by half of the quantum cloud offerings, but the concrete realizations differ:
For example, when using hybrid jobs for AWS Braket or sessions for Azure Quantum, once the job is started, it gets priority over regular jobs but not over other hybrid jobs or sessions which are currently running.
In contrast, IBMQ sessions provide dedicated and exclusive access to the quantum computer once the session is started and until a certain time window is exceeded.
Finally, the reservation of a time slot for exclusive access to a quantum computer is supported by the majority of quantum cloud offerings.
However, this typically incurs high costs.\looseness=-1

\noindent
\textbf{Batch Processing:}
The majority of the evaluated quantum cloud offerings supports the execution of multiple quantum circuits as a batch. 
However, some quantum cloud offerings, e.g., Quantum Inspire, limit the number of jobs submitted simultaneously, preventing the submission of a large number of circuits.
While batch processing can be emulated by providing a function that accepts a batch of circuits and then sequentially submits them to the cloud offering, this increases latency.

\noindent
\textbf{Circuit Cutting:}
Circuit cutting is currently not supported by any of the investigated quantum cloud offerings.
However, corresponding functionalities are included in some SDKs, such as Qiskit and Pennylane.
To ease the usage of circuit cutting techniques, IBMQ plans to automatically apply circuit cutting and the combination of execution results in their systems in 2025.\looseness=-1

\noindent
\textbf{Error Handling:}
All quantum cloud offerings except two enable directly executing error mitigation techniques without the need to manually apply them. 
However, some quantum cloud offerings only provide this feature for a subset of the available quantum computers, e.g., quantum computers from IonQ but not from Rigetti in Azure Quantum.
Although quantum error correction is currently supported by none of the quantum cloud offerings, some of the roadmaps of these offerings include this feature.
For example, IBMQ plans to demonstrate error correction in 2026 and include it in their systems by 2029 and Quandela Cloud intends to support error correction in 2027.\looseness=-1

\begin{figure*}[htb!]
	\centering
	\includegraphics[width=0.865\textwidth]{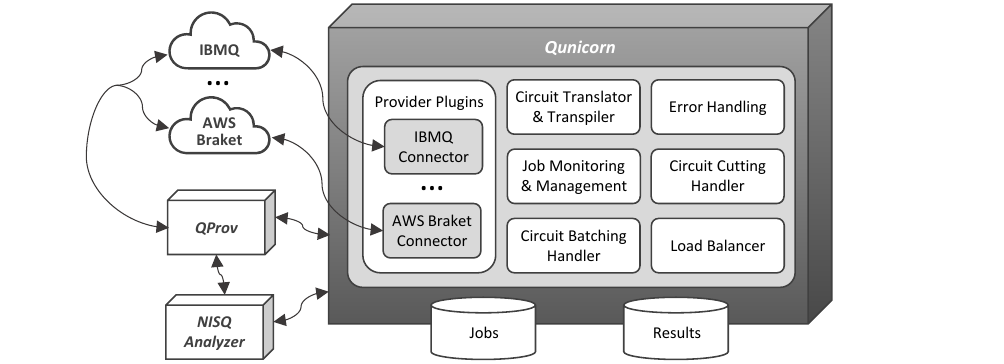}
	\caption{Overview of the system architecture for the unified execution across heterogeneous quantum cloud offerings}
	\label{fig:qunicorn}
\end{figure*}

\noindent
\textbf{Hybrid Runtime:}
Support for hybrid quantum applications based on a dedicated runtime is provided by AWS Braket, Azure Quantum, and IBMQ.
Some of the runtimes, e.g., from AWS Braket, only enable a combined deployment and prioritized access, while others, such as IBMQ, optimize the latency when switching between quantum and classical hardware.\looseness=-1

\noindent
\textbf{Integrated Development Environment:}
Circuit composers are rarely used by the evaluated quantum cloud offerings as they only enable modeling small quantum circuits which can be used for educational aspects but not to exploit the capabilities of today's quantum computers.
In contrast, online code editors are provided by the majority of quantum cloud offerings as they ease the required setup and \mbox{enable the creation and execution of larger quantum circuits.}

\noindent
\textbf{Pricing Models:}
Most of the evaluated quantum cloud offerings either use a time-based or circuit-based pricing model.
Only AWS Braket and Azure Quantum utilize both pricing models, providing more flexibility to the users.
The subscription-based pricing model is currently only supported by Azure Quantum.
Thereby, users get a fixed execution time for a monthly fee.
Quantum Inspire currently does not use any of the evaluated pricing models, as it can be used without costs.\looseness=-1

\noindent
\textbf{Trial Access:}
Trial access to quantum computers is made available in different forms:
For example, Alice \& Bob provides one hour of execution time per month for free.
In contrast, other quantum cloud offerings, such as Azure Quantum, add an initial amount of credits to newly created accounts, which can be used to execute quantum circuits.
Other quantum cloud offerings, such as AWS Braket, only support trial access to simulators.
Furthermore, trial access might be restricted, e.g., for academic purposes, which is the case for some quantum cloud offerings, such as IonQ.

\section{Unification Middleware for \\Quantum Cloud Offerings}
\label{sec:qunicorn}
\noindent
In this section, we tackle our second research question by introducing ~\textit{Qunicorn}, a unification middleware enabling the utilization of various quantum cloud offerings via a single, unified API.
First, the system architecture is introduced, and then we showcase the prototypical implementation realizing it.\looseness=-1

\subsection{System Architecture}
\label{subsec:arch}
\noindent
An overview of the system architecture realizing our approach for unifying the execution of quantum circuits is shown in \cref{fig:qunicorn}.
The system architecture comprises three components:
The Qunicorn middleware enables the unified execution of quantum circuits, \textit{QProv}~\cite{Weder2021_QuantumProvenance} collects provenance information about the different quantum cloud offerings, and the \textit{NISQ Analyzer}~\cite{Salm2020_NISQAnalyzer} is used for selecting a suitable quantum \mbox{computer} based on the circuit that shall be executed. 
Furthermore, Qunicorn connects to a variety of quantum cloud offerings, e.g., IBMQ or AWS Braket, as depicted on the left of \Cref{fig:qunicorn}.

Qunicorn consists of seven components:
The \textit{provider plugins} handle the communication with the different quantum cloud offerings.
Each quantum cloud offering is connected via a dedicated connector that wraps the API of the respective quantum cloud offering and translates the request and result formats.
This may also require the translation of the quantum circuits if their format is not supported by the target quantum computer, which is done by the \textit{circuit translator \& transpiler}.
Further, it compiles the quantum circuits for the topology of the utilized quantum computer.
The negative effect of errors is reduced by the \textit{error handling} component.
It provides access to different error handling methods, supporting automated error handling for all connected quantum cloud offerings.
If the quantum cloud offering to use provides dedicated error handling functionalities, the user can define if this functionality or the error handling capabilities of Qunicorn should be used.
The \textit{job monitoring \& management} component provides unified monitoring and management access, i.e., all submitted quantum circuits can be handled via a single interface independently of the quantum cloud offering that is actually used.
Typical monitoring and management operations are, e.g., checking the queue status, aborting a job, or retrieving execution results.
To enable the execution of quantum circuits that are too large for a quantum computer, the \textit{circuit cutting handler} can automatically cut quantum circuits into multiple sub-circuits, which are subsequently executed.
Similar to error handling, the user can specify if circuit cutting should be performed by Qunicorn or the quantum cloud offering once it is supported by one of the connected quantum cloud offerings.
The execution results of these sub-circuits are then combined to compute the result of the original circuit.
Since not all quantum cloud offerings support the batch processing of quantum circuits, a \textit{circuit batching handler} is used.
It accepts multiple quantum circuits and processes them as a single batch if this is supported by the utilized quantum cloud offering.
Otherwise, it creates a virtual batch that can be monitored by the user and adds the quantum circuits sequentially to the queue of the quantum computer.
The \textit{load balancer} enables the distribution of quantum circuits across different quantum computers, to speed up the execution of large amounts of quantum circuits. 
Finally, two databases are employed to store all available information about \textit{jobs} and execution \textit{results}.

The provenance system QProv~\cite{Weder2021_QuantumProvenance} supports the collection and storage of provenance data in a persistent database.
It contains an automated provenance collector that regularly collects data about the characteristics of quantum computers provided through different quantum cloud offerings.
This data can be used to improve the understandability and reproducibility of results, mitigate occurring errors, and select suitable quantum computers for a given circuit, based on the characteristics of the utilized quantum computer at the time of execution.

The NISQ Analyzer~\cite{Salm2020_NISQAnalyzer} enables selecting suitable quantum computers based on a given quantum circuit and the characteristics of the available quantum computers.
Qunicorn forwards quantum circuits to the NISQ Analyzer, receives the information about the selected quantum computer, and executes it via the provider plugin of the respective quantum cloud offering.\looseness=-1

\subsection{Prototypical Validation}
\noindent
To validate the practical feasibility of the introduced system architecture, we realize it as a prototypical implementation and utilize it for solving three exemplary application scenarios from the quantum computing domain.
The source code of all used components as well as detailed instructions describing how to set up our prototype and use it for the discussed application scenarios are publicly available on GitHub~\cite{USTUTT_qunicorn}.\looseness=-1

Qunicorn is implemented using \textit{Flask}, a \textit{Python} web framework. 
It provides a REST API that enables the execution of quantum circuits on a variety of different quantum cloud offerings through a single, unified endpoint.
The job management is handled asynchronously utilizing \textit{Redis} as a broker and \textit{Celery} as a queuing system. 
By implementing the Qunicorn functionalities as separate, loosely coupled modules, they are independently scalable and can easily be extended in the future. \looseness=-1

In the first scenario, we solve the 3-SAT problem using \textit{Grover`s algorithm}~\cite{grover1996fast}.
The implementation was done using the Braket SDK and the quantum circuit should be executed using the IBMQ quantum cloud offering.
Thus, Qunicorn translates the received quantum circuit and sends it to the offering using the Qiskit SDK.
Furthermore, error mitigation should be applied to reduce the impact of noise on the execution results.
This is enabled using the Qunicorn API and the user receives the mitigated execution results without further required actions.
The second application utilizes VQE~\cite{tilly2022VQE} to compute eigenvalues which is required in various scenarios, e.g., drug discovery and material design~\cite{hasan2022chemistry}.
As VQE requires executing multiple quantum circuits per iteration, the batch processing feature of Qunicorn is utilized.
Finally, in the third scenario, the quantum approximate optimization algorithm~\cite{farhi2014quantum} is used to solve the maximum cut problem.
Thereby, circuit cutting is applied using Qunicorn to reduce the size of the executed circuits, and thus, reduce errors during circuit execution.\looseness=-1

\section{Discussion \& Threats to Validity}
\label{sec:discussion}
\noindent
In this section, we discuss threats to validity regarding the performed gray literature review comparing quantum cloud offerings and limitations of the presented unification middleware.\looseness=-1

To avoid an insufficiently representative set of websites in the initial search of the gray literature review, a multi-phase search strategy using three well-established search engines was used.
A potential threat is that some websites are missed as they do not explicitly mention one of the utilized search terms or are not indexed by one of the used search engines.
This risk was reduced by using a query string that consists only of high-level, generic keywords, reducing the number of false negatives.
Another risk is the removal of relevant websites and quantum cloud offerings during the next phases, which is mitigated by employing two independent researchers and utilizing a deterministic set of inclusion and exclusion criteria.
These inclusion and exclusion criteria were initially validated in a set of trial runs and refined based on the obtained results.\looseness=-1

To enable the reproducibility and verifiability of our gray literature review and the quantum cloud offering comparison, we carefully documented all performed steps of the search and selection process.
The corresponding data is made publicly available~\cite{USTUTT_replication}.
However, the different utilized search engines do not provide suitable APIs that can be freely accessed to automate the search and extraction process.
Therefore, the obtained results require significant post-processing efforts, forming a potential problem for the reproducibility of our study.
Furthermore, some of the investigated quantum cloud offerings only provide the required information after logging into the corresponding offering.
Hence, reproducing the results of the quantum cloud offering comparison requires creating an account for each of the eight evaluated quantum cloud offerings.\looseness=-1

The prototypical implementation of Qunicorn currently supports two different quantum cloud offerings, namely AWS Braket and IBMQ.
This limits the applicability of the unification middleware if another quantum cloud offering is required.
However, Qunicorn provides a plugin-based architecture, facilitating the integration of additional quantum cloud offerings.\looseness=-1

Typically, quantum cloud offerings require passing a token to identify a user and execute quantum circuits.
Qunicorn currently accepts user tokens over its API when receiving an execution request, which are then forwarded to the quantum cloud offerings for execution.
However, to enable an automated quantum hardware selection as discussed in \cref{subsec:arch} the token for all potentially used quantum cloud offerings must be provided by the user.
Therefore, in the future, Qunicorn can be extended to store user tokens for different quantum cloud offerings.
Another option would be to utilize shared tokens \mbox{and support a dedicated pricing model as part of Qunicorn.}

\section{Related Work}
\label{sec:rw}
\noindent
Various works summarize existing quantum cloud offerings and present some of their features.
Nguyen~et~al.~\cite{nguyen2024_quantumcloudreview} review current trends in quantum cloud computing, including quantum applications for the cloud and quantum cloud offerings.
Thereby, they also discuss some of the features of quantum cloud offerings, e.g., the number of qubits of the largest quantum computer.
Moguel~et~al.~\cite{moguel2024development} compare the quantum cloud offerings AWS Braket, Google Quantum AI, IBMQ, and Azure Quantum.
For this, they describe some of their features and present their software stacks.
However, both works examine a smaller feature set and evaluate less quantum cloud offerings.
Vietz~et~al.~\cite{Vietz2021_QuantumSoftwareEngineeringChallenges} present a study exploring various challenges when realizing hybrid quantum applications in the cloud by analyzing existing quantum cloud offerings and their features.
Similarly, Obst~et~al.~\cite{Obst2023_ComparingQuantumOfferings} investigate how the execution of a typical application scenario differs when utilizing different quantum cloud offerings and document the lessons learned.
However, they do not systematically identify the quantum cloud offerings and analyze their feature set.

Several works focus on the execution of quantum circuits via different heterogeneous quantum cloud offerings.
The \textit{NISQ Analyzer}~\cite{Salm2020_NISQAnalyzer} introduced in \cref{subsec:arch}
also supports translating the quantum circuit after completed hardware selection into another format if this is required for the execution on the selected quantum computer.
In contrast to Qunicorn, it does not provide additional features, such as cutting quantum circuits or providing unified observability.
Beisel~et~al.~\cite{Beisel2023_Quokka} introduce \textit{Quokka}, a service ecosystem for quantum applications.
Quokka includes a quantum circuit execution service, which enables executing quantum circuits using different quantum cloud offerings.
However, it only handles the communication with the offerings and requires the quantum circuits to be in a format supported by the respective offering.
Garcia-Alonso~et~al.~\cite{garcia2021quantum} propose a \textit{Quantum API Gateway} to abstract the details required to execute a quantum circuit by automatically selecting a suitable quantum computer and handling the communication with the quantum cloud offering.
The Quantum API Gateway currently only enables the execution using different quantum computers available via a single quantum cloud offering, namely AWS Braket.
Nguyen~et~al.~\cite{Nguyen2022_QFaaS} present a framework for developing quantum functions which can be deployed and invoked as a service following the Function-as-a-Service~(FaaS) paradigm.
In contrast, Qunicorn focuses on a more lightweight approach and directly executes quantum circuits, instead of deploying quantum programs as independent functions.
Leymann~et~al.~\cite{Leymann2019_SharingQuantumSoftware} and Falkenthal~et~al.~\cite{falkenthal2024planqk} introduce a platform for sharing quantum software.
While this community platform enables the execution of quantum circuits via different quantum cloud offerings, its focus is sharing knowledge and implementations for quantum algorithms, as well as commercializing these implementations.
Stirbu~et~al.~\cite{stirbu2024_Qubernetes} present \textit{Qubernetes}, a tool for deploying hybrid quantum applications.
Although they propose a unified execution model, which focuses on hybrid quantum applications, Qunicorn addresses the unification on a lower abstraction level, i.e., the execution of quantum circuits.
Giortamis~et~al.~\cite{giortamis2024orchestratingquantumcloudenvironments} introduce a cloud orchestrator for hybrid quantum applications called \textit{Qonductor}.
Qonductor provides a unified API that enables the estimation of required resources for a given hybrid quantum application and the automated deployment of the contained quantum and classical programs.

\section{Conclusion \& Future Work}
\label{sec:conclusion}
\noindent
Quantum computers enable solving various problems faster or more energy-efficiently than classical computers.
However, quantum computers and the quantum cloud offerings used to access them are very heterogeneous, e.g., the supported programming languages or pricing models.
Thus, the selection of a suitable quantum cloud offering for a quantum application is difficult and often leads to a vendor lock-in due to the used technologies to realize the application.
In this paper, we performed an analysis of various quantum cloud offerings and evaluated their features to support developers in choosing a suitable quantum cloud offering.
Further, we presented an architecture and a prototypical implementation of a unification middleware, which enables the automatic translation of quantum circuits and results for the target quantum cloud offering.
To demonstrate the practical feasibility, we realized three exemplary application scenarios using our middleware.

In future work, we plan to evaluate the presented middleware in a user study with partners from both academia and industry.
The results will be used to increase the usability of our prototype and to identify additional required features.
Furthermore, we will integrate additional quantum cloud offerings in Qunicorn and incorporate newly developed \mbox{features provided by the supported quantum cloud offerings.}

\section*{Acknowledgements}
This work was partially funded by the BMWK projects \textit{SeQuenC}~(01MQ22009B) and \textit{EniQmA}~(01MQ22007B).

Furthermore, the authors would like to thank Julian Obscht for his great support in implementing the presented use cases.

\footnotesize
\bibliographystyle{IEEEtranN}
\bibliography{bibliography, masterbib}

All links were last followed on \today.

\end{document}